\documentclass[prl,twocolumn,showpacs,nofootinbib]{revtex4}

\usepackage{amsmath}    
\usepackage{graphicx}   
\usepackage{verbatim}   
\usepackage{color}      
\usepackage{subfigure}  
\usepackage{hyperref}   

\begin{document}

\title{Resonant enhancement of a single attosecond pulse in a gas medium by a time-delayed
control field}
\date{\today}
\author{Wei-Chun Chu and C.~D.~Lin}
\affiliation{J. R. Macdonald Laboratory, Department of Physics, Kansas State University,
Manhattan, Kansas 66506, USA}
\email{wcchu@phys.ksu.edu}

\begin{abstract}
An optical coherent control scheme has been proposed and theoretically investigated where an
extreme ultraviolet single attosecond pulse (SAP) propagates through a dense helium gas dressed
by a time-delayed femtosecond laser pulse. The laser pulse couples the $2s2p(^1P)$ and
$2s^2(^1S)$ autoionizing states when the SAP excites the $2s2p$ state. After going through the
gas, the spectral and temporal profiles of the SAP are strongly distorted. A narrowed but
enhanced spike in the spectrum shows up for specific intensities and time delays of the laser,
which exemplifies the control of a broadband photon wave packet by an ultrashort dressing field
for the first time. We analyze the photon and electron dynamics and determine the dressing
condition that maximizes this enhancement. The result demonstrates new possibilities of
attosecond optical control.
\end{abstract}

\maketitle



Single attosecond pulses (SAPs) were produced in the laboratory for the first time in
2001~\cite{hentschel}. The synchronized pair of an SAP and an infrared (IR) pulse in the so
called ``pump-probe'' scheme has become an indispensable tool over the past decade to
time-resolve the electron motion in quantum processes such as photoionization~\cite{kitzler},
Auger emission~\cite{drescher}, tunneling~\cite{uiberacker}, molecular
dissociation~\cite{kelkensberg}, and autoionization~\cite{gilbertson}. In most of these
experiments, to retrieve the dynamics, the SAP photoionizes the atoms or molecules, and the
``probe'' IR pulse ``streaks'' the photoelectrons with different time delays between the
pulses. The streaking model is based on electron emission to a structureless continuum energy
band. In reality, there are abundant excited states in atoms and molecules, which can be
strongly coupled by the IR field. This coupling scheme is not different from that in the
electromagnetically induced transparency (EIT) effect~\cite{marangos, fleischhauer-rmp}, which
has been studied extensively since first discovered in the 1990's~\cite{harris, boller} using
two laser pulses. In fact, the EIT effect has been studied recently in extreme ultraviolet
(XUV)~\cite{loh, gaarde, tarana, pfeiffer} and X-ray~\cite{buth, glover} energies in tens to
hundreds of femtoseconds.


Using light to control quantum systems has long been of great interest not only for the
fundamental understanding of light-matter interactions, but also for applications in quantum
information~\cite{lukin-rmp}, light source generation~\cite{sansone}, ultracold
atoms~\cite{wieman}, and precision spectroscopy~\cite{hall, hansch}. The EIT effect has been a
powerhouse for optical coherent control where photoabsorption of light in matter in certain
energy range can be reduced by a precisely configured dressing laser. As a prominent
application in quantum information, photon storage is achieved by EIT where the refraction
index is tuned to transfer information between photonic and atomic states~\cite{fleischhauer,
liu, phillips}. To achieve efficient coherent transitions, this type of optical control is
carried out in nanosecond to microsecond timescales and under perfect resonance condition
between long-living quantum states.


Very lately, after the color-shifting of a light pulse was proposed~\cite{notomi} and
realized~\cite{preble} using crystals, an alternative approach using the EIT technique with
time-delayed pulses was reported~\cite{ignesti10}. The same approach is also capable of tuning
the bandwidth of a propagating light~\cite{ignesti11}. In the time domain, a transient
``transmission gain'' has been observed in nanoseconds when the strong dressing field is
rapidly switched on or off~\cite{chen, echaniz, greentree}. All of these efforts not only
elucidate the EIT mechanism, but also promote the level of manipulation of light from the
traditional EIT effect to higher degrees of control. However, these techniques are established
for narrowband lasers and unable to treat the dynamics initiated by a broadband SAP in
attosecond timescales.


The present theoretical work aims to control an SAP going through a laser-dressed gaseous
medium by utilizing the state-of-the-art ultrafast technology. We simulate the propagation of
an 1-fs XUV pulse (bandwidth 1.8~eV) in a helium gas where the $2s2p(^1P)$ and $2s^2(^1S)$
autoionizing states (AISs) are strongly coupled by a 9-fs laser pulse. The dynamics of the
coupled system is controlled by three comparable timescales, given by the autoionization, the
population transfer between the two AISs, and the durations of the two light pulses. By tuning
the laser intensity and the time delay between the pulses to specific values, a sharp peak,
with a 50\% increase from the incident signal, shows up in the XUV transmission spectrum at the
$2s2p$ resonance. Compared to the transparency window in EIT, it is effectively a photon
emission window controlled by the ultrashort dressing laser. The enhanced peak is very
localized in energy ($<100$~meV) and in the time delay ($<3$~fs), demonstrating a high-level
control of a photon wave packet.



\begin{figure}[htbp]
\centering
\includegraphics{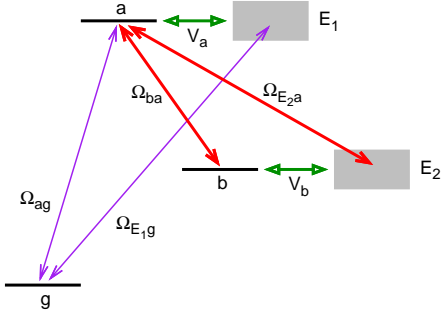}
\caption{(Color online) Schematic diagram of a three-level atom in two light fields. The arrows
represent coupling between the states, labeled by the transition amplitudes. The green arrows
are autoionizations, the thin purple arrows are XUV couplings, and the thick red arrows are
laser couplings. Only first-order transitions are considered.}
\label{fig:scheme}
\end{figure}

Consider a three-level atom, where the top two levels $|a\rangle$ and $|b\rangle$ are AISs,
associated with the background continuum $|E_1\rangle$ and $|E_2\rangle$, respectively. An XUV
(labeled by $X$) pulse couples the ground state $|g\rangle$ to the $|a\rangle$-$|E_1\rangle$
resonance, and a laser (labeled by $L$) pulse couples the two resonances, where the radiation
fields are treated classically~\cite{fontana}. The coupling scheme can be either ladder- or
$\Lambda$-type. For the near-resonance coupling with weak to moderately strong field
intensities, the total wavefunction is described by~\cite{lambropoulos, bachau, madsen,
themelis}
\begin{eqnarray}
|\Psi(t)\rangle &= e^{-iE_gt} c_g(t) |g\rangle \nonumber\\
&+ e^{-iE_Xt} \left[ c_a(t) |a\rangle + \int{c_{E_1}(t) |E_1\rangle dE_1} \right] \nonumber\\
&+ e^{-iE_Lt} \left[ c_b(t) |b\rangle + \int{c_{E_2}(t) |E_2\rangle dE_2} \right],
\label{eq:Psi}
\end{eqnarray}
where $E_X\equiv E_g+\omega_X$, $E_L\equiv E_g+\omega_X+\omega_L$, and the $c(t)$'s are
smoothly varying coefficients. The three-level system is schematically plotted in
\ref{fig:scheme}. For a given set of atomic levels and external fields, the total wavefunction
$|\Psi(t)\rangle$ is uniquely calculated~\cite{chu11}.

By convention, we assume the form of light pulses as $E(t)=F(t)e^{i\omega t}+c.c.$. The
envelope $F(t)$ is complex in general, and it is initially a sine-square type real function.
All the fast oscillating terms are factored out in the numerical calculation. The dipole moment
of the atom is~\cite{chu12}
\begin{equation}
\mu(t) = u_X(t) e^{i\omega_Xt} + u_L(t) e^{i\omega_Lt} + c.c., \label{eq:mu}
\end{equation}
where
\begin{eqnarray}
u_X(t) &\equiv \left( D_{ag} + i\pi V_a D_{E_1g} \right) c_a^*(t) c_g(t) \nonumber\\
&- i\pi F_X(t) \left| D_{E_1g} c_g(t) \right|^2, \label{eq:uX}
\end{eqnarray}
and $u_L(t)$ corresponds to laser frequency, which is not in our concern. The matrix elements
$D_{ij}\equiv \langle i|D|j\rangle$ and $V_a \equiv \langle E_1|H|a \rangle$ are real values by
employing standing waves for the continuum basis $|E_1\rangle$ and $|E_2\rangle$.


For a given set of XUV and laser pulses, the wavefunction $|\Psi(t)\rangle$ and the dipole
response $\mu(t)$ are adequate to describe the electron emission and photoabsorption in a
dilute gas~\cite{chu11, chu12}. However, in a dense gas, the fields are altered by the response
of the atomic medium. By pairing a weak XUV and a moderately strong laser that are different by
two orders of magnitude in the intensity, we assume that the laser propagates without
modification in the medium. The ionization is weak so the plasma effect is disregarded. The
spontaneous decay is much slower than the AIS decay and dismissed in the model.

By redefining the time coordinate in the moving frame $t'=t-z/c$ where $z$ is along the
traveling direction of the light, the Maxwell equation is written as
\begin{equation}
\frac{\partial E(z,t')}{\partial z} = -\frac{\rho}{c\epsilon_0} \frac{\partial
\mu(z,t')}{\partial t'}, \label{eq:E-mu}
\end{equation}
where $\rho$ is the density of the gas. The dependence in the transverse direction has been
removed for loosely focused lights. Factoring out the carrier oscillation terms, the
propagation of the XUV field is described by
\begin{equation}
\frac{\partial F_X(z,t')}{\partial z} = -\frac{\rho}{c\epsilon_0} \left[ \frac{\partial
u_X(z,t')}{\partial t'} + i\omega_X u_X(z,t') \right]. \label{eq:F-u}
\end{equation}
Putting together (\ref{eq:uX}) and (\ref{eq:F-u}), one can integrate iteratively to obtain the
dipole envelope $u_X(z,t')$ and the XUV envelope $F_X(z,t')$ as functions of $t'$ at each
spatial point $z$ until the pulse exits the medium, which is then defined as the transmitted
XUV pulse.


The medium in concern is a 2-mm thick gas of noninteracting helium atoms. The pressure and the
temperature are 25~torr and 300~K respectively (density $\rho = 8\times 10^{17}$~cm$^{-3}$). An
XUV pulse (SAP) and a time-delayed laser pulse are colinearly propagated and linearly polarized
in the same direction. The 1.8-eV bandwidth of the XUV pulse (central frequency
$\omega_X=60.15$~eV, duration in full width at half maximum $\tau_X=1$~fs, and peak intensity
$I_X=10^{10}$~W/cm$^2$) is much broader than the 37-meV $2s2p$ resonance. The laser pulse
($\lambda_L=540$~nm, $\tau_L=9$~fs) resonantly couples $2s2p$ and $2s^2$. For convenience, the
peak laser intensities ($I_L$) are expressed in terms of $I_0 \equiv 10^{12}$~W/cm$^2$. The
time delay $t_0$ is defined as the time between the two pulse peaks, and it is positive when
the XUV peak comes first.

\begin{figure}[htbp]
\centering
\includegraphics{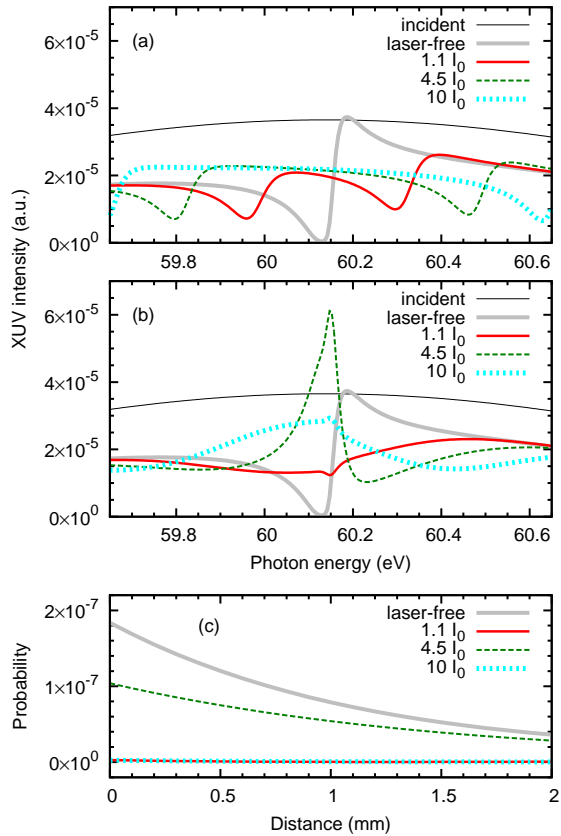}
\caption{(Color online) (a) XUV spectra near the $2s2p$ resonance with a 1-ps overlapping
($t_0=0$) laser. (b) Same as (a) but the laser duration is 9~fs. Various peak laser intensities
are applied. (c) The $2s2p$ bound state population at the end of the laser pulse versus the
distance inside the gas.} \label{fig:spectrum}
\end{figure}

To reproduce the EIT condition as a reference, we apply a 1-ps laser pulse which overlaps
temporally with the SAP and calculate the XUV spectra. As seen in \ref{fig:spectrum}(a),
without the laser, the spectrum simply displays the $2s2p$ Fano lineshape~\cite{fano}. When the
laser intensity increases, the Fano peak splits into two with increasing separations. Note that
these intensities correspond to linearly increasing field strengths. The split peaks are the
exact reproduction of the Autler-Townes doublet~\cite{autler}, where the peak separation is
equal to the Rabi frequency. Although our ``probe'' in this EIT-like setup is,
unconventionally, an SAP, as long as the laser is long in terms of the coupling strength, the
dressed-state interpretation is valid, and the Autler-Townes phenomenon is recovered. Note that
the split lineshapes are broadened due to the 120-meV resonance width of $2s^2$.

Next, we shorten the laser duration to $\tau_L=9$~fs and keep the overlapping condition
($t_0=0$). The XUV spectra are plotted in \ref{fig:spectrum}(b). For $I_L=1.1I_0$, the Fano
lineshape that appears in the laser-free case is washed out, and the spectrum is almost flat,
which indicates that the ionization path through the ``bound'' $2s2p$ vanishes. For
$I_L=4.5I_0$, the central region of the spectrum turns to an upside-down mirror image of the
original Fano lineshape, implying that the path through $2s2p$ has gone through a phase shift
of $\pi$, or equivalently, the Fano $q$-parameter changes sign after one Rabi cycle. What is
more surprising is that in 60.0-60.2~eV, the transmission signal is actually higher than the
incident signal, i.e., the XUV emission is produced at the resonance by the particular coupling
condition. However, by further increasing the intensity to $I_L=10I_0$, the emission signal
drops down, and the enhancement feature disappears.

The disappearance of the $2s2p$ peak in the XUV spectrum for certain laser intensities is
clarified by the $2s2p$ bound state population $|c_a(t)|^2$. Its value at the moment the laser
pulse ends (12.4~fs) is plotted in \ref{fig:spectrum}(c) along the propagation in $z$, before
it further decays toward the end of autoionization. Under the laser-free condition, the SAP
intensity drops exponentially as it is absorbed along $z$, and thus the excitation of $2s2p$
decreases accordingly. For $I_L=1.1I_0$, $4.5I_0$, and $10I_0$, the Rabi oscillation after the
SAP runs for 0.5, 1, and 1.5 cycles, respectively, where the $2s2p$ population is either dried
up (half-integer cycles) or highly populated (integer cycles). For $I_L=4.5I_0$, the $2s2p$
residue keeps decaying to the resonance peak, while for $I_L=1.1I_0$ and $10I_0$, the spectra
turn to be quite flat.

\begin{figure}[htbp]
\centering
\includegraphics{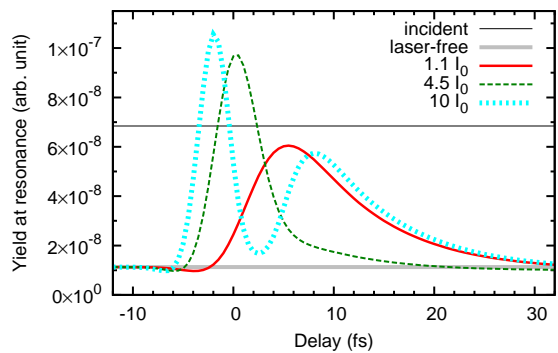}
\caption{(Color online) Total XUV yields covering a 50-meV window centered at the $2s2p$
resonance against the time delay, for different laser intensities.}
\label{fig:yield}
\end{figure}

The comparison in \ref{fig:spectrum}(b) shows that the resonant absorption turns to emission
only for a specific laser intensity. To investigate the trend further, \ref{fig:yield} plots
the XUV signal in a 50-meV energy window centered at $2s2p$ (60.15~eV) against the time delay.
This spectral resolution is easily reachable in typical spectrometers. Without the laser, the
incident and transmitted signals are $7\times10^{-8}$ and $1\times10^{-8}$ respectively, i.e.,
85\% of the light in this window is absorbed by the medium. With the presence of the coupling
laser, the signal yield builds up with time delay, reaching a maximum, and then decreases
again. As the intensity increases to $I_L=4.5I_0$, the buildup starts faster, and the maximum
yield at $t_0=0$ is about 40\% higher than the incident signal. For $I_L=10I_0$, the maximum
enhancement increases to about 50\% and shifts to $t_0=-2$~fs. These extraordinary oscillatory
features for different laser intensities could be revealed in an actual XUV-plus-IR type
measurement.

To fully understand the mechanism behind this signal oscillation, the dynamics in the time
domain is analyzed. Under laser-free condition, the SAP prompts the $2s2p$ state and generates
dipole oscillation, which then radiates XUV before the $2s2p$ decays. The transmitted light
consists of the leading SAP (``head'') and the longer trailing part (``tail'') due to the
emission from the decaying $2s2p$. Their coherent sum results in the original Fano line shape
seen in \ref{fig:spectrum}. The head and the tail partially cancel each other in the temporally
overlapping region due to the phase shift of the dipole radiation, which is analogous to the
study carried out for a rectangular pulse propagating in a two-level atomic
system~\cite{shakhmuratov} as a form of transient nutation~\cite{hocker}. In the presence of
laser coupling, the Rabi oscillation brings the electrons back and forth between the two AISs
after the SAP. Every Rabi cycle that brings the electrons back to $2s2p$ adds a phase shift of
$\pi$ to the tail. Transmission enhancement occurs when there is just one Rabi cycle to flip
the dip part (the minimum) to a maximum due to the additional $\pi$ phase added to the
interference. Note that regardless of the laser intensity, the magnitude of the dipole
oscillation initiated by the SAP always declines with time. In other words, the earlier the
Rabi cycle is accomplished, the stronger the enhancement appears. This aspect is seen by the
peak positions and heights of all three curves in \ref{fig:yield}. This prediction complements
the earlier experiment using an attosecond pulse train to enhance the high harmonic
signals~\cite{gademann}, which was based on the scheme first conducted to study wave packet
interference~\cite{johnsson}. They were related to our study but in a different nature.

As expressed by the propagation in (\ref{eq:F-u}), the XUV pulse is shaped by the electric
dipole modulated by the 2.3-eV laser field. The process is a form of spectral redistribution of
the XUV pulse caused by the oscillation between the AISs. To illustrate the conservation of the
total XUV signal in this process, we integrate each curve in \ref{fig:spectrum}(b) over an 1-eV
range centered at the resonance which covers the essential reshaping region of interest.
Without the laser, nearly 40\% of the XUV photons are absorbed as the pulse goes through the
medium. When the laser is turned on, for all three intensities, the yield decreases only
slightly from the laser-free case, by at most about 17\%. This extra loss is dissipated via the
autoionization of $2s^2$, mediated by the coupling laser.

\begin{figure}[htbp]
\centering
\includegraphics{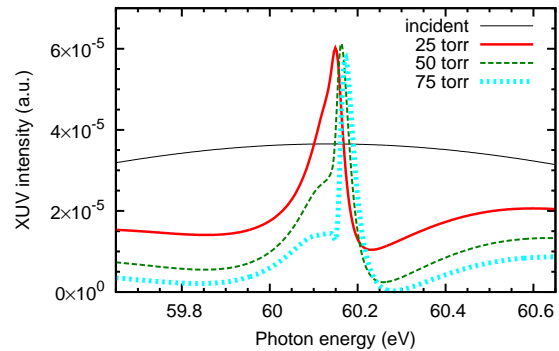}
\caption{(Color online) Transmitted XUV spectra for the gas pressures of 25, 50, and 75~torr.
The laser condition is $\lambda_L=540$~nm, $\tau_L=9$~fs, $I_L=4.5$~TW/cm$^2$, and $t_0=0$.}
\label{fig:pressure}
\end{figure}

Light traveling in a linear medium attenuates with a constant absorption rate proportional to
the medium density, as stated by Beer's law~\cite{ishimaru}. However, in a strongly coupled
medium, this law does not hold anymore, especially for high densities. The XUV transmission
spectra for $I_L=4.5I_0$, $t_0=0$, and gas pressures $P=25$, 50, and 75~torr are displayed in
\ref{fig:pressure}. The figure shows that the background continuum part is weakened as the
pressure rises, but the enhanced peak, while distorted and shifted, remarkably maintains the
same height. For $P=75$~torr, the highest transmitted light signal is about 20 times larger
than the lowest part in the background. The stability of the enhancement peak is due to the
balance between two processes: the light is dragged from the pulse head backward in time ($t'$)
to the tail, while the tail keeps being absorbed; both processes are accelerated by higher gas
densities.


In conclusion, we have simulated an 1-fs XUV pulse propagating through a dense gas of
three-level autoionizing helium atoms coupled by a time-delayed 9-fs laser. By adjusting the
coupling laser to preferred conditions, an enhancement appears at the resonance energy in the
XUV transmission spectrum, demonstrating up to a drastic 50\% increase in the signal count. The
transmitted XUV pulse is composed by the attosecond head part and the longer tail part in the
time domain, where the phase of the tail can be sectionally shifted by the laser, forming
special interference patterns in the frequency domain. This remarkable feature relies on the
cutting edge of ultrafast technologies--both the XUV and laser pulses are considerably shorter
than the 17-fs decay lifetime of $2s2p$--to clearly define the timing of the wave packet
dynamics. We find that while higher density obstructs the light surrounding the resonance, the
enhancement peak can sustain, so that it spikes up more apparently. Our simulation undertakes
reasonable experimental setup and conditions. The unique measurable features and the underlying
dynamics could invoke the profound possibilities in the attosecond nonlinear optics.

\begin{acknowledgments}
This work is supported in part by Chemical Sciences, Geosciences and Biosciences Division,
Office of Basic Energy Sciences, Office of Science, U.S. Department of Energy.
\end{acknowledgments}

\end{document}